# Disentangling pre-transitional fluctuations in metallic $VO_2$

Javier del Valle[1,2*], Carl Willem Rischau[3], Artem Korshunov[4], David Ambrosi-Jalón[1], Aitana Tarazaga Martín-Luengo[1,2], Ibraheem Yousef[6], Sara A. Lopez-Paz[5], Shany Neyshtadt-Ronel[7], Reetendra Singh[7], Abhishek Rakshit[7], Stefano Gariglio[3], Alexei Bosak[4] and Yoav Kalcheim[7]

[1]Department of Physics, University of Oviedo, C/ Federico García Lorca 18, 33007 Oviedo, Spain
[2]Center of Research on Nanomaterials and Nanotechnology, CINN (CSIC-Universidad de Oviedo), El Entrego 33940, Spain
[3]Department of Quantum Matter Physics, University of Geneva, 24 Quai Ernest-Ansermet, 1211 Geneva, Switzerland
[4]European Synchrotron Radiation Facility, BP 220, 38043 Grenoble Cedex, France
[5]Department of Chemistry, University of Copenhagen, Universitetsparken 5, 2100 Copenhagen, Denmark
[6]MIRAS Beamline BL01, ALBA-CELLS Synchrotron, Cerdanyola del Vallès, 08209, Barcelona, Spain
[7]Department of Material Science and Engineering, Technion - Israel Institute of Technology, Haifa 32000, Israel

*Corresponding author: javier.delvalle@uniovi.es

## Abstract

$VO_2$ features concomitant structural and metal-insulator transitions. This poses a challenge for understanding the underlying mechanism: is the transition triggered by a structural or by an electronic instability? The two scenarios are expected to produce very different pre-transitional fluctuations above $T_C$. By combining magnetic susceptibility, IR reflectivity and X-ray diffuse scattering measurements, we observe that metallic $VO_2$ features strong electronic and structural fluctuations towards the insulating monoclinic phase. By measuring resonant diffuse X-ray scattering across the Vanadium K-edge, we search for a potential decoupling between electronic and structural ordering in these fluctuations, finding no evidence of it. While our results do not completely rule out pure electronic fluctuations, they constrain them, favoring the interpretation that the $VO_2$ metal-insulator transition is triggered by a structural instability. Our work offers a novel approach to solve similar problems in other strongly correlated systems.

## Introduction

Many strongly correlated materials feature phase transitions in which both the electronic and structural order parameters change simultaneously. Concomitant electronic and structural transitions are observed in systems as diverse as Kagome metals [1], colossal magnetoresistance manganites [2], charge density wave dichalcogenides [3,4] or metal-insulator transition (MIT) oxides [5–7]. In all these cases, understanding the driving force triggering the transition is complicated due to coincident changes in spin, charge, orbital, and lattice ordering. Two scenarios are generally proposed: the transition is induced either by a lattice instability or by an electronic instability [8–10]. Figure 1 shows a schematic representation of both scenarios in terms of free energy, an analysis inspired by recent work by Georgescu *et al.* [8].

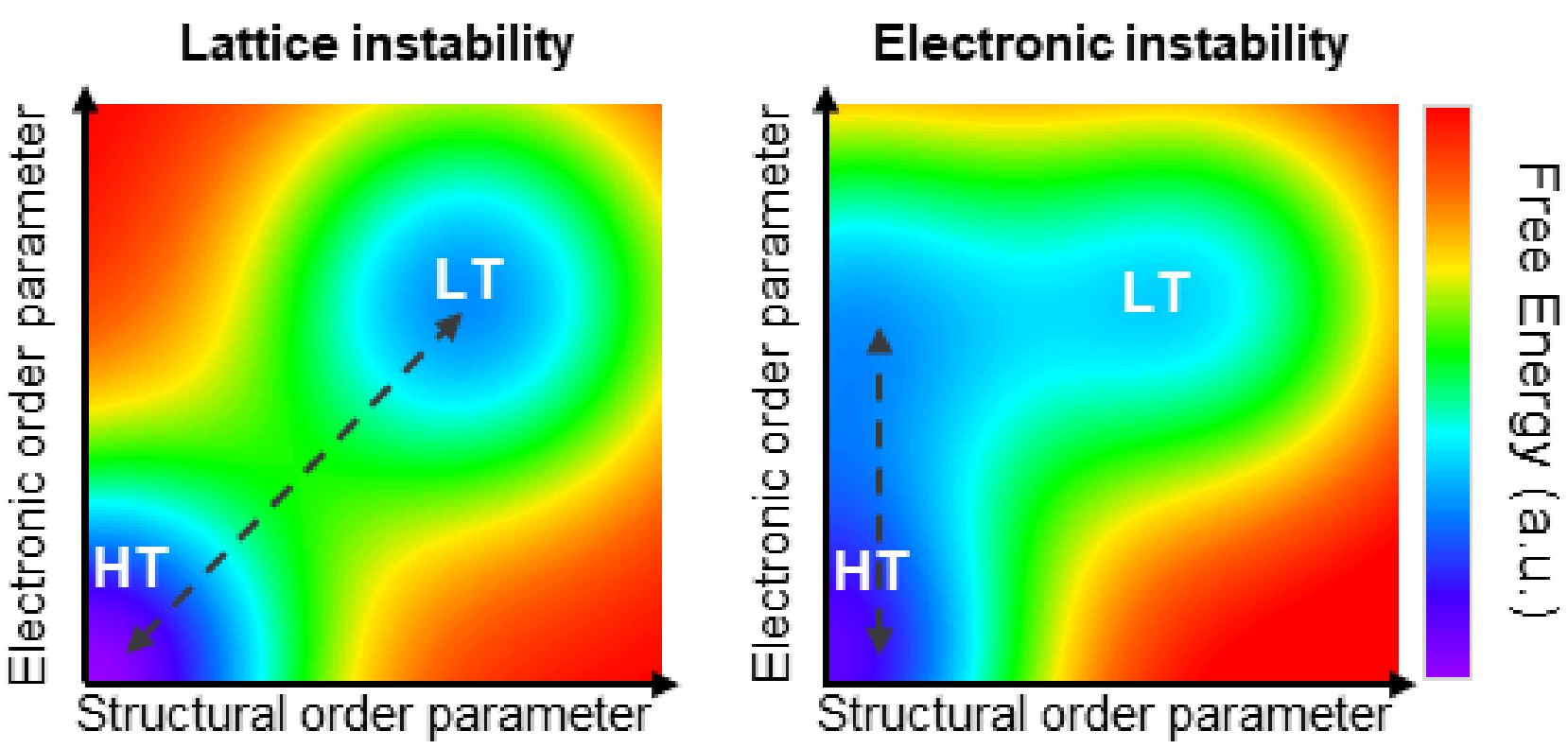

Figure 1. Schematic representation of the two scenarios proposed to describe concomitant electronic and structural transitions. The free energy is plotted as a function of two order parameters: structural and electronic. The two local minima correspond to the high (HT) and low temperature (LT) phases. In this specific example, $T>T_C$, so the undistorted HT phase is the global minimum and the LT phase is metastable. Dashed arrows show the fluctuations from the HT phase expected for the two scenarios.

In the structural instability picture, both the ion cores and valence electrons reorganize jointly to reduce the overall energy of the system [8]. One example of such instability would be the Peierls transition. In the electronic instability picture, electronic interactions alone reorder the configuration of valence electrons, without the need for a lattice deformation [11,12]. A classic example is the metal-insulator transition of the Mott-Hubbard model [13]. In this latter case, the lattice could still deform, but only as a consequence of the valence electron redistribution. Since the lattice follows the valence electronic charge in timescales as fast as 100 fs [14], it is very challenging to experimentally differentiate between the two scenarios. This egg-or-chicken problem has resulted in a longstanding debate in the condensed matter community.

$VO_2$ is a paradigmatic example of this dilemma, undergoing a phase transition from a rutile paramagnetic metal into a monoclinic diamagnetic insulator at 340 K [7,15]. These are usually termed the R and M1 phases, respectively. In the M1 phase, all vanadium atoms dimerize, *d* electrons form singlet pairs, and the resulting large interdimer hopping renders the system insulating. Goodenough made the first theoretical description of the transition, proposing the bonding-antibonding splitting induced by dimerization as the key element opening the insulating gap [16]. This is schematically described in Figure S1a,b [17]. Later calculations showed that while this lattice-driven picture is conceptually correct, it is necessary to include strong electron-electron correlations for the insulating gap to open, making the MIT a correlations-assisted Peierls transition [9,10,18–20]. Such a picture would still be described by the left panel in Figure 1.

The discovery of a second monoclinic insulating phase (M2 phase, see phase diagram in Fig. S1c), where only half of the vanadium atoms dimerize [21], challenged the naïve lattice-driven picture and suggested that electron-electron interactions might dominate [22–24]. DFT/DMFT (Density Functional Theory/Dynamic Mean Field Theory) calculations placed the R phase near a Mott instability [22], and further experiments showed this phase to be a strongly correlated metal on the brink of localization. This was evidenced by a resistivity above the Ioffe-Regel limit [25], Planckian dissipation [26], the presence of a pseudogap [27,28] and exotic thermal transport signatures [29]. More recent experiments have even proposed that the MIT and structural transitions of $VO_2$ and the related compound $V_2O_3$ can take place independently [12,30–33], although these results have subsequently been challenged, both in the equilibrium [34–36] and ultrafast regimes [37]. Despite decades of research, a consensus on what triggers the MIT in $VO_2$ is still lacking, exemplifying the broader challenge of understanding coupled electronic and structural phenomena in correlated materials.

A potential way to tackle this problem is to study pre-transitional fluctuations in the metallic phase just above $T_C$, which indirectly probe the free-energy landscape illustrated in Fig. 1. The two proposed mechanisms should produce qualitatively different fluctuation patterns. In a lattice-driven transition, electronic and lattice distortions are expected to take place together. In contrast, in an electronically driven scenario the free-energy surface permits substantial electronic fluctuations with only minimal lattice distortion (dashed arrows in Fig. 1). Above $T_C$, such electronic fluctuations can exist on fs timescales, where the lattice cannot respond efficiently. Upon reaching $T_C$, the electronic instability condenses, valence-charge distortions become static, and the lattice subsequently reorganizes on phonon timescales (hundreds of fs).

Such electronic fluctuations could, in principle, be of two types: symmetry-preserving or symmetry-breaking**.** In the symmetry-preserving case, $d_{\parallel}$ electrons (see Fig. S1a,b) would transiently localize on individual V atoms, reducing interatomic overlap, suppressing hopping, and locally opening a gap—while maintaining the overall rutile symmetry. As discussed later, this scenario appears unlikely given the strong structural fluctuations observed in the metallic R phase. In the symmetry-breaking case, the $d_{\parallel}/\pi^*$ orbitals would distort, breaking local inversion symmetry by asymmetrically shifting valence charge density from the V sites towards one of the V-V interatomic bond regions. Recent DFT+U+V [38,39] calculations show a difference of up to 0.4 electrons in interatomic charge density between short and long V–V bonds in the insulating M1 phase. The charge depletion along some of the bonds reduces hopping, locally opening a gap. It is conceivable that such valence charge displacements could precede, and eventually drive structural fluctuations in the metallic state. Within this framework, the electronic and structural order parameters in Fig. 1 correspond to the interatomic charge density and the V–V dimer bond length, respectively.

Here we study pre-transitional fluctuations in $VO_2$ single crystals using a variety of probes. First, we show that metallic $VO_2$ exhibits pronounced electronic and structural fluctuations toward the insulating monoclinic phase. This is revealed by a combination of magnetic susceptibility, infrared reflectivity, and X-ray diffuse scattering measurements. To test whether electronic and lattice degrees of freedom remain coupled within these fluctuations, we performed resonant diffuse X-ray scattering across the vanadium K-edge. Because the scattering time at the V K-edge (~few fs [40]) is faster than typical electronic or lattice timescales, any transient decoupling between charge and structure would manifest in the diffraction pattern. These measurements, presented in the second half of the paper, show no evidence of purely electronic fluctuations. However, given the relatively small resonant cross section, their presence cannot be entirely excluded. By analyzing the expected energy dependence of thermal diffuse scattering and the experimental detection limits, we estimate that any purely electronic fluctuations with coherence lengths below ~2 u.c. or above a few u.c. are unlikely. Our results largely constrain the type of pre-transitional fluctuations present in $VO_2$ and favor the interpretation that the transition is triggered by a structural instability. Moreover, our findings establish a methodology to experimentally address similar problems in other strongly correlated materials beyond MITs.

**Pre-transitional electronic fluctuations**

We fabricated $VO_2$ single crystals using the self-flux method, annealing $V_2O_5$ powder for 24 hours at 1000º C under an Argon gas flow [41]. High quality single crystals with lengths up to several mm can be obtained this way.

Figure 2a shows the resistivity vs temperature ($T$) of one of these crystals, which undergoes a five orders of magnitude resistivity change at the MIT. The inset in Figure 2a presents synchrotron-based infrared reflectance measurements in the R phase, performed at the MIRAS beamline at the ALBA synchrotron. The spectral feature near 0.2 eV, located well above the phonon spectrum, corresponds to the pseudogap previously reported by Qazilbash et *al.* [27] and Huffman et *al.* [28]. The presence of a pseudogap suggests that in the metallic state, conduction electrons are not completely itinerant, but are partially localized. Similar pseudogaps have been reported in the metallic state of related compound $V_2O_3$ [42].

Figure 2b shows magnetic susceptibility ($\chi$) as a function of temperature (300 K - 1000 K). The diamagnetic to paramagnetic transition is readily visible close to 340 K. Surprisingly, the paramagnetic $\chi$ has a strong temperature dependence. This is unexpected for metals, usually dominated by temperature-independent Pauli paramagnetism [43]. The inset in Figure 2b shows $\chi^{-1}$ vs $T$. Although not quite linear in temperature, a behavior reminiscent of Curie paramagnetism can be appreciated, again indicating that conduction electrons are partially localized. The slope of the $\chi^{-1}$ ($T$) curve is not constant but becomes flatter as the MIT is approached. A crude extrapolation results in a negative temperature intercept for $\chi^{-1}$=0, consistent with antiferromagnetic correlations [43] that would favor the formation of spin singlets. Together, the IR reflectivity and the magnetic susceptibility results suggest the presence of localized conduction electrons and the emergence of pre-transitional spin singlets in the R phase of $VO_2$.

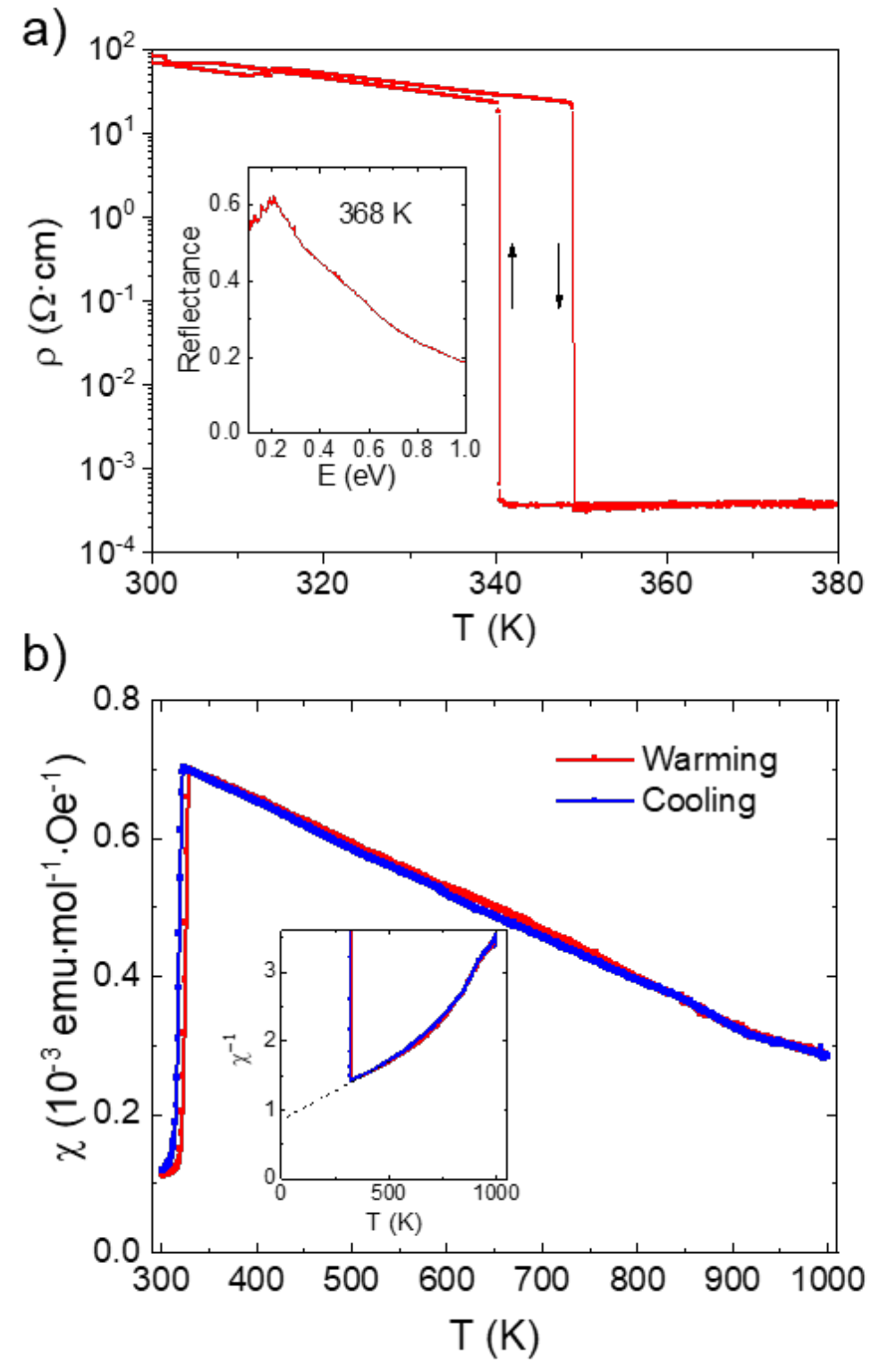

Figure 2. a) Resistivity vs $T$. Inset: Infrared reflectance vs photon energy measured in the R phase at T=368 K. b) $\chi$ vs $T$. Inset: $\chi^{-1}$ vs $T$. The dashed line is an extrapolation obtained by a linear fitting of the 350 K – 550 K range.

**Pre-transitional structural fluctuations**

Metallic $VO_2$ features a rutile structure consisting of chains of edge-sharing $VO_6$ octahedra interconnected via the apical oxygen, as shown in Figure S2 [17]. The strongest structural correlations are thus contained within the $(\pm1\pm1\ 0)_R$ planes [44]. Figure 3a shows the structure of the R, M2, and M1 phases in the $(1\text{-}1\ 0)_R$ plane. As pointed out by Pouget *et al*. [7], the M2 structure can be derived from the R structure by considering 1-dimensional, chain-like shifts of vanadium positions along the $[11\pm1]_R$ and $[-1\text{-}1\pm1]_R$ directions. These are represented with black arrows. The interchain interaction results in the M2 structure: dimerization and buckling along $[001]_R$, as schematically depicted in the central panel of Figure 3a. The M1 structure can be further derived from the M2 phase by also including chain-like shifts contained in the perpendicular plane, along the $[1\text{-}1\pm1]_R$ and $[-11\pm1]_R$ directions [7,44].

Pre-transitional, one-dimensional $[\pm1\pm1\pm1]_R$ displacements can be strikingly visualized thanks to diffuse X-ray scattering, as reported before [7,45]. Chain-like disorder is expected to manifest as 2-dimensional planes of diffuse scattering in reciprocal space. Figure 3b shows total X-ray scattering measurements performed in ID28 at the ESRF using a photon energy of 15.8 keV. Reciprocal space slices parallel to the (HK0) plane, as well as the (HHL) plane are displayed. In addition to the Bragg peaks from the average rutile structure, intense streaks of diffuse scattering can be seen. These correspond to intercepts of diffuse scattering planes oriented perpendicularly to the $[\pm1\pm1\pm1]_R$ directions. As previously shown [41,45,46], these planes do not arise from static disorder but from thermal diffuse scattering (TDS) coming from transverse acoustic flat phonon bands that become softer as the MIT is approached. They pass through the *R* and *M* high symmetry points (see Figure S3 [17] for a schematic representation of the R phase Brillouin zone).

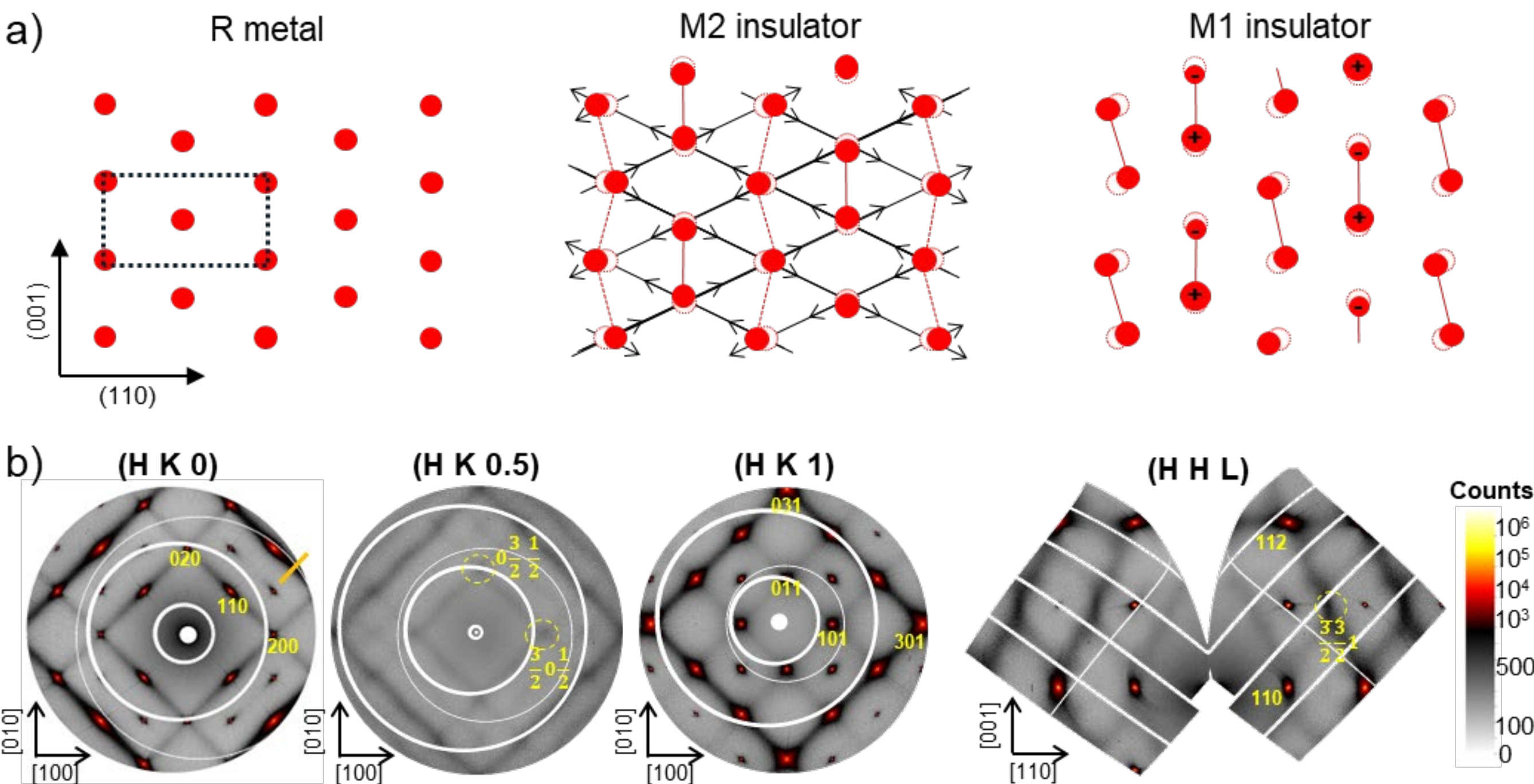

Figure 3. a) Schematic representation of the structure of the R, M2, and M1 phases, viewed along the [1 -1 0] direction. Solid red circles represent Vanadium atoms. Dashed empty circles show the original position of the vanadium atoms in the R-phase. Black arrows show the alternating chain-like displacements along the $[11\pm1]_R$ and $[-1-1\pm1]_R$ directions. Solid circles with - and + signs indicate displacements in and out of the page. The dotted rectangle shows the basic unit cell in the R phase b) X-ray total scattering in different slices of the reciprocal space, measured in the R phase at 350 K: (HK0), (HK0.5), (HK1), and (HHL). Several Bragg peaks and high symmetry *M* and *R* points are outlined in yellow.

The observation of structural fluctuations does not, by itself, rule out an electronically driven transition as depicted in Fig. 1. If electronic fluctuations locally break inversion symmetry and become sufficiently slow, they can couple to the lattice, softening the phonon dispersion and giving rise to the observed TDS—in other words, purely electronic fluctuations could induce structural ones. In this scenario, the coherence volume of charge fluctuations should exceed that of the structural distortions, which is about 2 u.c., as inferred from the width of the diffuse scattering planes (~0.35 Å$^{-1}$). Conversely, the presence of strong structural distortions effectively rules out symmetry-preserving charge fluctuations since they would not drive the lattice toward lower symmetry.

**Resonant diffuse X-ray scattering**

The results in Figures 2 and 3 demonstrate that, in the metallic R phase of $VO_2$, both electronic and structural degrees of freedom exhibit pronounced pre-transitional fluctuations toward the insulating M1/M2 phases. However, the techniques used cannot determine whether valence-electron fluctuations follow a different spatial pattern than lattice distortions, or whether they remain rigidly coupled.

Resonant elastic X-ray scattering (REXS) offers a direct way to address this question [47,48]. At resonance, the X-ray scattering cross section of valence electrons is strongly enhanced, making REXS a powerful probe of electronic ordering [49–53], dynamic correlations [54]. In the present case, we are interested in fluctuations involving V 3d electrons. While the V L-edge would probe these states directly through dipole-allowed 2p–3d transitions, it does not provide sufficient photon momentum to access the regions of reciprocal space where the diffuse scattering planes are observed. We therefore probe the V 3d states at the K edge (5.47 keV), focusing on the pre-edge region corresponding to the 1s → 3d transition.

The V K pre-edge has been shown to be highly sensitive to the oxidation state and the MIT [55], and REXS at this energy has successfully been applied to detect magnetic and orbital ordering in $V_2O_3$ [50]. Although the 1s → 3d transition is formally quadrupolar (E2–E2) and thus weak in centrosymmetric environments, local inversion-symmetry breaking enables p–d orbital mixing, allowing much stronger dipolar (E1–E1) contributions to the pre-edge signal [55,56]. Because the electronic fluctuations considered here involve symmetry-breaking redistributions of valence charge, they are expected to produce a measurable enhancement of the diffuse scattering intensity near the pre-edge, making REXS a sensitive probe of such fluctuations.

REXS measurements were carried out at the P09 beamline of PETRA III (DESY) with an energy resolution of 0.7 eV. A pyrolithic graphite analyzer was used to remove the fluorescent background and analyze the scattered light polarization. TDS features shown in Fig. 3 can be resolved with our resonant scattering setup, with typical signals in the order of 500-5000 counts/s. This is demonstrated in Fig. S4, which displays (H,H,0) line scans through a diffuse plane at several photon energies. The fluorescence signal across the K edge, which shows a pronounced pre-edge, is presented in Fig. 4a. Figures 4b and 4c plot the scattered X-ray intensity, collected in a $\sigma$–$\sigma'$ polarization configuration, as a function of photon energy at the (110) Bragg peak and at the *M* and *R* high-symmetry points lying in the diffuse-scattering planes. These measurements were performed with the incoming light polarization nearly parallel to the c-axis -and $d_{//}$ orbital-. Corresponding data for the remaining high-symmetry points of the rutile lattice are shown in Fig. S5. Figure S6 displays equivalent measurements at the M and R points for incoming light polarized nearly perpendicular to the c-axis. Except for the (110) Bragg peak, all intensities were corrected for absorption (see Supplementary Material [17] and Fig. S7). All data were collected at 380 K, in the metallic phase.

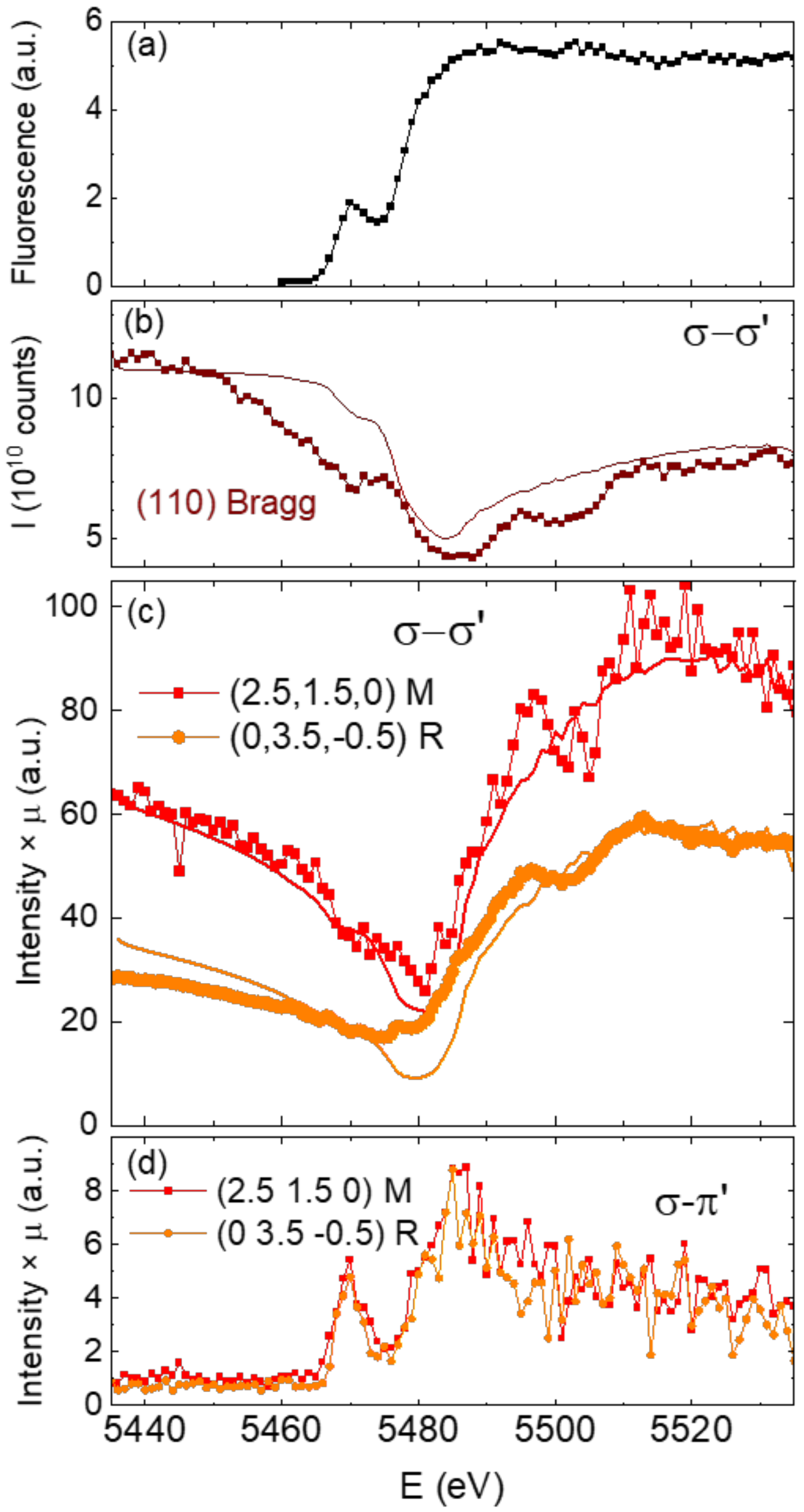

Figure 4. a) Sample fluorescence measured using the V Kβ line. b) X-ray intensity vs photon energy for the (110) Bragg reflection. Solid dark red line shows the expected behavior (see supplementary information [17]). c) Absorption-corrected $\sigma$–$\sigma'$ X-ray intensity vs photon energy for the $M$ and $R$ high symmetry points, contained within the diffuse scattering planes. Red and orange solid lines show the expected behavior, proportional to $\left|f(\vec{Q},E)\right|^2$(see supplementary information [17]). d) Absorption-corrected $\sigma$–$\pi'$ X-ray intensity vs photon energy for the $M$ and $R$ high symmetry points. T=380 K.

## Discussion

We first discuss measurements in a $\sigma$–$\sigma'$ configuration. In our samples, TDS is the main source of scattering away from Bragg peaks. Low-energy acoustic phonons, dominated by Vanadium atoms, contribute most to TDS [45]. We therefore expect that, across the K edge, TDS intensity will be roughly proportional to the atomic form factor of $V^{4+}$ ions: $f(\vec{Q},E) = f_0(\vec{Q}) + f'(E) + if''(E)$, where $\vec{Q}$ is the scattering vector and $E$ the photon energy. The solid curves in Fig. 4c show the calculated TDS energy dependence for the $M$ and $R$ points, showing good agreement with the experimental data. Details of the calculation are provided in Supplementary Material [17] and Fig. S8. All measured high-symmetry points and azimuths display the same dependence (Fig. S5 and S6), confirming that diffuse scattering across the K edge originates from phonon-induced disorder. Importantly, no enhancement near the absorption edge or pre-edge—typical of electronic ordering—is observed at the (111) planes or at any high-symmetry point of the R-phase structure.

Measurements in the $\sigma$–$\pi'$ channel look very different. Because TDS is not expected to contribute to this polarization configuration, the background remains close to zero before the absorption edge. Two clear resonances appear at the pre-edge and main-edge energies. As shown in Figs. S5 and S6, these resonances do not depend on the sample orientation or the position in reciprocal space. This lack of momentum and azimuth dependence indicates that the signal does not originate from coherent fluctuations involving multiple atoms, but instead from local, incoherent processes. We attribute these features to RIXS-like contributions -such as *d-d* or charge transfer excitations- that enter our measurement due to the relatively large energy bandwidth (~15 eV) of the crystal analyzer. Such processes can rotate the polarization [57], giving rise to intensity in the $\sigma$–$\pi'$ channel. Similar features are likely present in the unrotated $\sigma$–$\sigma'$ channel but are obscured there by the strong TDS background. Fig. S9 shows intensity versus energy at the M point measured in the unrotated $\pi$–$\pi'$ channel with $2\theta = 91.6^\circ$ to suppress TDS contributions; the pre-edge resonance is clearly visible despite the large background.

d-electron ordering would produce a sharp pre-edge resonance with distinct reciprocal-space structure, which we do not observe. However, this absence does not exclude electronic fluctuations: very short-range correlations would yield a broad, weak feature that could be hidden beneath the strong TDS background. In the Supplementary Information [17], we elaborate a simple model to estimate the expected resonant intensity as a function of the fluctuation coherence volume, normalized to the TDS signal. Caveats and limitations are also discussed there. The results, shown in Fig. 5 for the M and R points, indicate that electronic fluctuations with coherence volumes greater than a few u.c. would be detectable. We must note that, because our measurements probe only high-symmetry wavevectors, electronic fluctuations that do not involve unit-cell doubling cannot be excluded. But since the structural fluctuations do intersect two high symmetry points, this scenario is unlikely.

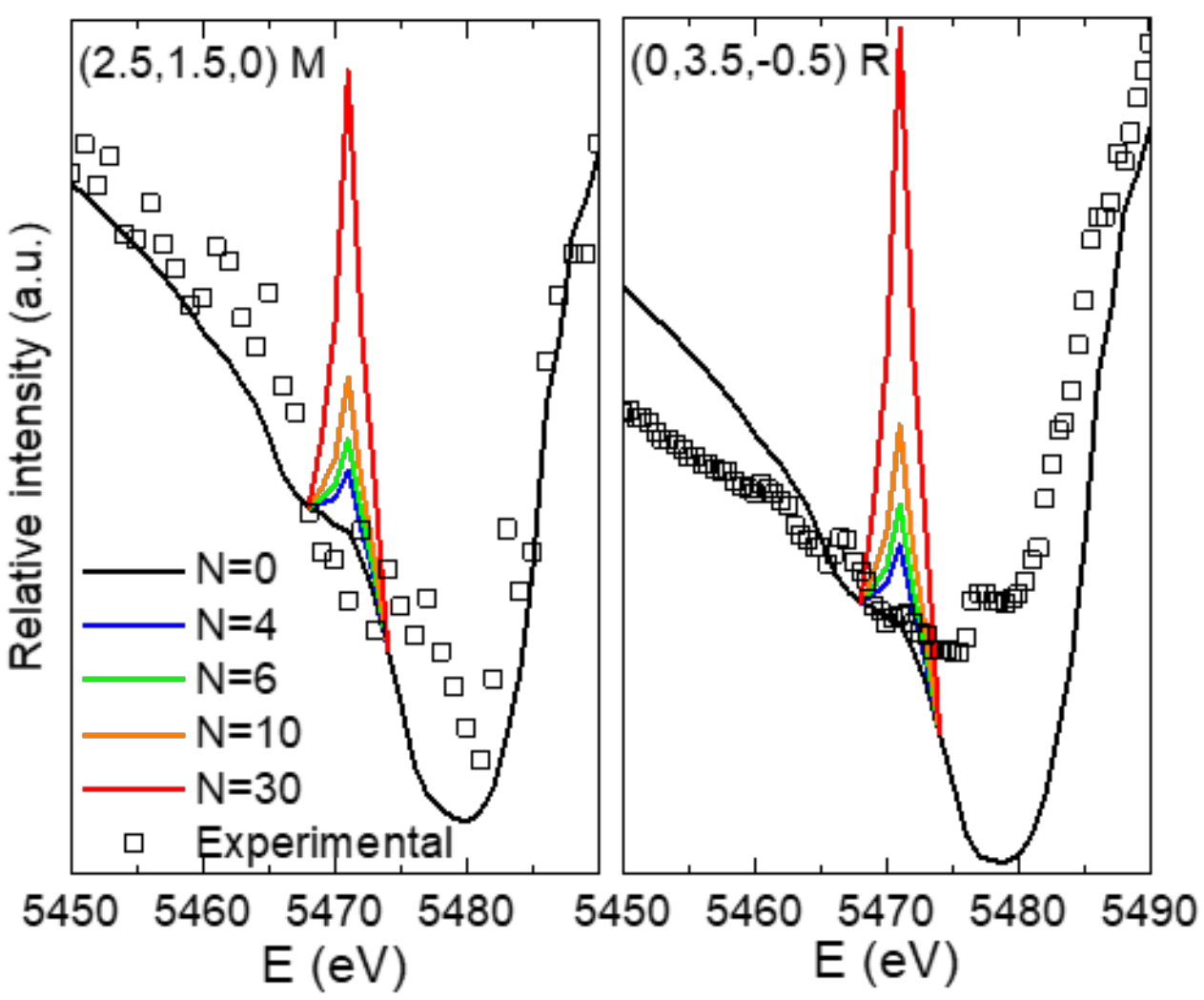

Figure 5. Expected resonant signal for charge fluctuations of different coherent volumes, at the M (left) and R (right) points, as detailed in the supplementary material [17]. The solid black line is the expected energy dependence of the TDS.

Our work places clear constraints on the nature of potential electronic fluctuations, favoring the interpretation that the MIT is primarily driven by a structural instability. Such an incipient lattice instability can account not only for the observations in Figure 3 but also those in Figure 2. If the lattice is locally fluctuating into the M1/M2 structure, transient dimerization could take place, which would, in turn, localize pairs of carriers, induce antiferromagnetic correlations, and favor the formation of singlets. The measured pseudogap [27,28] may arise from the excitation of polarons, formed by the interaction between the deformed lattice and the trapped conduction electron, as is observed in some manganites [58,59]. Polarons have also been recently proposed to play a key role in the properties of the M1 phase and the triggering of the MIT [60–62].

Interestingly, pre-transitional dynamics are far from being limited to the proximity of the MIT. They set in hundreds of degrees above the transition and show no signs of criticality as the MIT approaches. The magnetic susceptibility varies slowly over a range of almost 700 K, and mid-IR reflectance measurements reveal no significant differences up to 573 K (Figure S10). Similarly, diffuse X-ray scattering patterns remain largely unchanged between 350 K and 500 K (Figure S11) [17]. These observations are in line with our recent report on the anomalous temperature dependence of phonon lifetimes in metallic $VO_2$ [41], and other studies highlighting the gradual softening of the R-phase lattice [45]. While our REXS measurements were performed only at T=380 K, we do not expect them to vary strongly with temperature.

The approach demonstrated here—probing for resonant enhancements in the diffuse scattering signal to identify potential electronic precursors—can be applied to address similar open questions in other materials. It is particularly promising for heavier transition-metal oxides, such as niobates, ruthenates, and iridates, where the metal L edge lies at sufficiently large photon momentum to sensitively probe TDS signatures.

## Conclusions

Using a range of experimental techniques, we characterized the metallic phase of $VO_2$ above the MIT, observing pronounced electronic and structural pre-transitional fluctuations toward the insulating state. Resonant diffuse X-ray scattering across the vanadium K-edge reveals no detectable decoupling between lattice and charge degrees of freedom in these fluctuations, supporting the structurally driven scenario for the MIT. Nonetheless, due to experimental sensitivity, purely electronic fluctuations with coherence lengths between 2 and ~10 unit cells cannot be excluded. Our approach establishes a general framework for addressing similar questions in other correlated materials, particularly heavier transition-metal oxides.

## Acknowledgements

We thank Giacomo Mazza, Lucas Korosec, Jean-Marc Triscone and Pablo Alonso González for fruitful discussions. We acknowledge DESY (Hamburg, Germany), a member of the Helmholtz Association HGF, for the provision of experimental facilities. Parts of this research were carried out at beamline P09 at PETRA III. Beamtime was allocated for proposal I-20230723 EC and we would like to thank Dr. Sonia Francoual for setting up the experiment and for useful discussions of the experimental data. We also thank the ESRF and ALBA synchrotrons for beamtime allocation and support during the measurements. This work was financed by the 'Proyecto de Generación de Conocimiento 2023' PID2023-147042NA-I00, funded by the Spanish MICIU/AEI/10.13039/501100011033, and by the European FEDER, UE. J.d.V was supported by a 'Ramón y Cajal' grant RYC-2021-030952-I from the Spanish MICIU/AEI. Part of the REXS measurements and sample growth were funded by the European Union's Horizon Europe research and innovation program under grant agreement (ERC, MOTTSWITCH, 101039986). Views and opinions expressed are however those of the authors only and do not necessarily reflect those of the European Union or the European Research Council Executive Agency. Neither the European Union nor the granting authority can be held responsible for them. D.A.-J. was supported by the Severo Ochoa program from the government of the Principality of Asturias (BP24-221). C.W.R. was supported by the U.S. Office of Naval Research through the NICOP Grant N62909-21-1-2028.

# Supplementary information

## 1. Experimental methods

Crystal growth, transport, and susceptibility measurements.

$VO_2$ single crystals were fabricated using the self-flux method, annealing $V_2O_5$ powder (99.95 %, Aldrich) for 24 hours at 1000º C in a horizontal tube furnace under an Argon gas flow. High-quality single crystals with lengths up to several mm can be obtained this way. Typically, 5 grams of $V_2O_5$ powder yield around 10 mg of hundreds of $VO_2$ crystals. Most of them are smaller than 100 μm, but many of them have lengths up to 1-2 mm and are large enough to be easily manipulated with tweezers. The crystals are needle-shaped, with the long axis oriented along $(001)_R$, and their side faces perpendicular to the $(110)_R$ direction

Transport measurements were done using a four-probe geometry in a Lakeshore TTPX probe station. Silver paste was used to contact the crystals. Magnetic susceptibility was measured using vibrating sample magnetometry in a EV9 magnetometer. 10.8 mg of $VO_2$ crystals were used. For susceptibility measurements, the crystals were partially crushed into smaller pieces to avoid alignment into a preferential direction, and they were kept under an $N_2$ atmosphere during the measurement to avoid oxidation at high temperatures.

Fourier-transform Infrared reflectivity (FTIR) measurements.

Synchrotron-based FTIR measurements were done in the BL01-MIRAS beamline at the ALBA Synchrotron light source (Barcelona, Spain). A Vector 70 spectrometer was coupled to a Hyperion 3000 microscope to perform SR-FTIR microspectroscopy measurements. The microscope is equipped with a liquid $N_2$-cooled HgCdTe 50 μm MCT detector and uses a 15X Schwarzschild objective (NA=0.40). The spectra were obtained in reflection mode, using a masking aperture size of 50 μm x 50 μm. The data were collected in the IR range (8500-800 cm-1) at a spectral resolution of 4 $cm^{-1}$ with 256 co-added scans per spectrum. A temperature control Linkam stage (THMS600) was used to warm the samples above room temperature. A gold reference was used to normalize the sample spectra.

Hard X-ray total scattering measurements.

The total X-ray scattering measurements shown in Figure 3 and S11 were collected on the ID28 beamline at the European Synchrotron Radiation Facility (ESRF).

A $VO_2$ crystal, glued with epoxy to the sample holder was measured in transmission geometry. A photon energy of 15.8 keV was used, and scattered radiation was collected with a Pilatus 2D detector. A $N_2$ heat blower was used to warm up the sample while preventing its oxidation.

Resonant Elastic X-ray scattering (REXS).

X-ray diffraction across the vanadium K-edge, shown in Figures 4,5, S4, S5, S6 and S9 was performed at the P09 beamline at PETRA III at DESY.

A $VO_2$ single crystal was glued onto the sample holder using silver paste. The sample holder was screwed into the cold finger of an ARS cryostat, isolated from the outside using a Beryllium dome and steel shroud, and then pumped down to $10^{-6}$ mbar for cryogenic operations. A Lakeshore 340 temperature controller was used to control the sample temperature.

Because of the low penetration depth at 5 keV and given the relatively large size of our crystals, a reflection Bragg geometry was used to measure the scattered intensity. Given the low momentum of photons at 5.48 keV and the reflection geometry, high 2θ angles were needed to reach the few available high symmetry points. Measurements were performed using an avalanche photodiode point detector to detect the weak features of diffuse scattering, and a pyrolithic graphite 001 analyzer for suppressing the fluorescence background and for polarization analysis. Energy resolution of the incidence beam was around 0.7 eV, and the energy window of the analyzer is ~ 15 eV.

## 2. Absorption correction

As the X ray beam travels through the sample, part of it gets absorbed. Absorption is greatly increased across the edge, which strongly distorts the shape of the intensity vs energy curves. To account for this, raw data corresponding to Figures 4c, 4d, S5 and S6 were corrected by multiplying the measured intensity $I$ by the linear absorption coefficient $\mu(E)$ [50]. $\mu(E)$ can be calculated from the fluorescence data. As shown by Eisebitt et al. [63], for a thick sample the measured fluorescence signal $F(E)$ will be given by:

$$F(E) = AI_0(E)\frac{\mu_K(E)}{\mu(E) + \mu(E_F)\frac{\sin\varpi}{\sin\beta}}$$

Where $E$ is the incoming photon energy, $E_F$ is the energy of fluorescent photons, $\varpi$ is the angle between incoming photons and the surface, $\beta$ is the angle between outcoming photons and the surface (see inset in Figure S7), $I_0(E)$ is the intensity of the incoming beam and $A$ is an experimental constant geometrical factor. $\mu(E) = \mu_K(E) + \mu_{other}(E)$ are the total absorption coefficient, the absorption coefficient due to the K edge and the absorption coefficient due to any other absorption edge (such as L or M edges), respectively. The value of $\mu_{other}(E) = 0.0267\ \mu m^{-1}$was extracted from the Henke tables [64] and taken constant in the energy range considered here. By measuring fluorescence at several incidence angles it is possible to extract $\mu(E)$, which is shown in Figure S7. The obtained $\mu(E)$ is in good agreement with previous reports on other vanadium oxide phases [50]. Once $\mu(E)$ is determined, the raw data is multiplied by it to account for the variation of the absorption coefficient across the edge.

This correction is only done for diffuse scattering data. The 110 Bragg peak shown in Figure 1b was not corrected for absorption. The reason for this is its high intensity (over $10^{11}$ counts), which is a sizeable fraction of the direct beam intensity. This is likely due to the high quality of the crystal (with a rocking curve width of ~0.01º) and it implies that diffraction, and not absorption, could be limiting how deep the X-ray beam is penetrating into the sample. The characteristic length over which the beam will be completely diffracted is known as extinction or Pendellösung length, and for a perfect crystal it is given by $\Lambda = V/r_e\lambda F_H$, where $V$ is the unit cell volume, $r_e$ is the classical electron radius and $F_H$ is the structure factor of the specific Bragg reflection [65]. For the 110 peak this length is around 1 μm, which is shorter than the absorption length ($1/\mu$) both below (37 μm) and above (5 μm) the K edge. This means that diffraction, and not absorption, is indeed the factor limiting how much volume of the sample that gets illuminated. Therefore, simply multiplying by $\mu$ leads to an overcorrection.

Instead, dynamical diffraction theory must be used to estimate the expected energy dependence of the Bragg peak. Following Authier [65], the solid line in Figure 4b was calculated in reflection geometry for an absorbing perfect crystal. We used the anomalous scattering coefficients calculated in the next section. We must note that in a lower quality crystal the diffracted energy will be limited by imperfections, and absorption effects dominate. In such case, the standard absorption correction can be used.

**3. Calculation of the anomalous scattering factors**

The atomic form factor $f(\vec{Q},E)$ of the $V^{4+}$ ion, for a particular scattering vector $\vec{Q}$ can be expressed as the sum of three terms [64]:

$$f(\vec{Q},E) = f_0(\vec{Q}) + f'(E) + if''(E)$$

Where $f_0(\vec{Q})$ is the usual Thomson term, and $f'(E)$ and $f''(E)$ are the real and imaginary parts of the anomalous scattering factor, respectively. $f_0(\vec{Q})$ is tabulated for most ions. $f''(E)$ is directly proportional to $\mu(E)$. It was calculated using $\mu(E)$ and the known values of $f''(E)$ just above and below the K edge, which were extracted from the Henke tables. $f'(E)$ was then derived from $f''(E)$ by applying a Kramers-Kronig transformation [64]:

$$f'(E) = \frac{2}{\pi}\int_0^\infty dE'\ E'f''(E')/(E^2 - E'^2)$$

The calculated $f'(E)$ and $f''(E)$ are shown in Figure S8. They are used to estimate the energy trends of measured intensities in Figures 4, 5, S5 and S6.

**4. Expected energy dependence of thermal diffuse scattering**

Thermal diffuse scattering (TDS) intensity $I(\vec{Q},E)$, coming from a specific phonon mode at $\vec{Q}$ with $\hbar\omega_0$ energy will be given by [66]:

$$I(\vec{Q},E) = \frac{\hbar N I_e}{2\omega_0} \cdot \coth\left(\frac{\hbar\omega_0}{2k_B T}\right) \cdot \left|\sum_s \frac{f_s(\vec{Q},E)}{\sqrt{m_s}} e^{-M_s} \cdot (\vec{Q}\cdot\vec{\varepsilon_s}) \cdot e^{i\vec{Q}\cdot\vec{d_s}}\right|^2$$

Where $N$ is the number of cells in the crystal, $T$ is the temperature, and the $s$ index runs through all atoms in the unit cell. $f_s(\vec{Q})$, $e^{-M_s}$, $\vec{\varepsilon_s}$ and $m_s$ are the atomic form factor, Debye-Waller factor, normalized displacement, and atomic mass of atom $s$, respectively. $E$ is the photon energy. The constant $I_e$ is proportional to the intensity scattered by a single electron, given by the Thomson scattering formula.

TDS in $VO_2$ is overwhelmingly dominated by low energy acoustic phonons consisting mainly of vanadium displacements. Taking this into account and using the experimental phonon dispersion from Budai et al., we estimate the TDS intensity per rutile unit cell. For the (3.5, 0, -0.5) R point it is expected to be:

$$\mathrm{I}(\vec{Q}) \approx 0.4 \cdot \left|f(\vec{Q},E)\right|^2 \cdot I_e$$

Around the K edge, the only parameter in the above equations that depends strongly on energy is $f(\vec{Q},E)$. Therefore, the energy dependence of TDS will be proportional to $\left|f(\vec{Q},E)\right|^2$. Solid lines in Figure 4c are proportional to this factor. Since $f_0(\vec{Q})$ depends on $\vec{Q}$, the curve shape will be slightly different for the different high symmetry points.

**4. Estimation of the resonant peak coming from potential electronic fluctuations**

Resonant scattering peaks originate from contrasts in the anomalous scattering factor, $\Delta f(\vec{Q},E)$, which develop at the pre-edge or main edge when inequivalent valence-electron configurations are present on otherwise identical atomic sites. In the present context, we consider hypothetical electronic fluctuations of *d* electrons that induce, at the pre-edge, a contrast $\Delta f_{pre-edge}(\vec{Q},E)$ between inequivalent V sites.

Because the microscopic nature of such fluctuations is not precisely known, a first-principles calculation of $\Delta f_{pre-edge}(\vec{Q},E)$ is not feasible. Computing the corresponding resonant scattering amplitude would require treating strong electronic correlations and substantial structural disorder on equal footing. In practice, this would entail extending dynamical mean-field theory (DMFT) to a structurally disordered system and explicitly evaluating the E1–E1, E1–E2, and E2–E2 transition matrix elements, for which no established theoretical framework currently exists. We therefore proceed by estimating the order of magnitude of the expected resonant contrast using experimentally constrained quantities. Before discussing the pre-edge signal of hypothetical electronic fluctuations, we will discuss the effect of the existing structural fluctuations.

The pre-edge features of rutile compounds $TiO_2$ and $VO_2$ are dominated by dipolar (E1-E1) transitions induced by p-d hybridization, with quadrupole (E2-E2) processes being considerably weaker [69,70]. In metallic $VO_2$ we get a pre-edge contribution to the anomalous scattering factor of $|f|_{Pre-edge} \approx 0.6$ electrons (Fig. S8). The pre-edge intensity of monoclinic $VO_2$ is just 1.5 times larger than that of the rutile phase [67,68,70]. Such small change is surprising, considering the V sites are locally centrosymmetric in the rutile phase and undergo a large off-centre distortion across the transition. The small difference is likely due to the intensity of pre-transitional structural fluctuations in the rutile phase ($\sqrt{\langle u_R \rangle^2}$ ≈0.18 Å) [45], which are comparable to the static V shift across the structural transition ($u_{M1,static}$ ≈0.23 Å). Such large dynamic displacements imply that V atoms are rattled far from the centre of the oxygen octahedron, thereby breaking local inversion symmetry and increasing the E1-E1 absorption channel.

The $VO_2$ pre-edge intensity can be decomposed into a baseline component due to covalency and quadrupolar terms, plus a symmetry-breaking-induced component: $f_{Pre-edge} = f_{Baseline} + f_{SB}$. Considering only first order corrections to the transition amplitude integral: $f_{SB} \propto u^2$ [71]. The ratio ${u_{M1,static}}^2/\langle u_R \rangle^2 \approx 1.6$ is close to the 1.5 increase in intensity across the transition. This suggests that pre-edge intensity in rutile $VO_2$ is dominated by $f_{SB}$, that is, by the symmetry-breaking- structural fluctuations. If atomic displacements follow a gaussian distribution, $u^2$ will follow a peak shaped distribution, with a peak width $\approx \langle u \rangle^2$. The anomalous scattering factor $f_{Pre-edge} \approx f_{SB} \propto u^2$, will roughly follow a similar statistical distribution, therefore giving a typical $\left|\Delta f_{pre-edge}\left(\vec{Q}, E\right)\right| \approx 0.6$ electrons.

We must note that we have made some important assumptions to get to this value. Specifically, we have supposed that: i) $f_{SB}$ for the static M1 distortions is approximately similar as $f_{SB}$ for rutile transient displacements; and ii) $f_{Baseline}$ is comparable in both phases, or at least much smaller than $f_{SB}$. We believe these assumptions are reasonable for a broad, order of magnitude estimate.

Let´s now consider the electronic instability scenario: the valence charge deforms first, and the ion cores eventually follow. In this scenario, valence orbitals would be even more asymmetrically distorted than in a simple structural fluctuation, and there would be a larger p-d mixing. This would yield a larger contrast at the pre-edge when compared to a naïve structural fluctuation where the orbitals deform as the nuclear coordinate evolves. For the moment, we will conservatively consider $\left|\Delta f_{pre-edge}\left(\vec{Q}, E\right)\right| \approx 0.6$ electrons both for structural and electronic fluctuations.

Once is $\left|\Delta f_{pre-edge}\left(\vec{Q}, E\right)\right|$ has been estimated, we can calculate the expected diffracted intensity. Let´s consider fluctuations (electronic or structural) with a short-range periodicity defined by a high-symmetry wavevector Q. The resulting resonant intensity per rutile unit cell can then be written as:

$$\mathrm{I}\left(\vec{Q}\right) \approx \left(N_f I_e e^{-2M_s}\right) \cdot \left|\Delta f_{pre-edge}\left(\vec{Q}, E\right)\right|^2$$

where $N_f$ is the fluctuation coherence volume, $e^{-M_s}$ is the Debye-Waller factor (~0.8 in our case [45]), and $I_e$ is the X-ray intensity emitted by a single electron, given by the Thomson scattering formula. We have assumed coherent addition of scattering amplitudes within the fluctuation volume.

Substituting the estimated $\left|\Delta f_{pre-edge}\left(\vec{Q}, E\right)\right|$ value into the above equation yields:

$$\mathrm{I}\left(\vec{Q}\right) \approx 0.3\left(N_f I_e\right)$$

We can now distinguish between two limiting scenarios. If the transition is purely structural, lattice fluctuations alone would be responsible for the pre-edge signal. The relevant coherence volume would be set by the extent of the structural distortions, which from Fig. 3 is $N_f \approx 2$, unit cells. This gives:

$$\mathrm{I}(\vec{Q}) \approx 0.6\, I_e$$

which is too weak to be experimentally detectable, as discussed in the main text. Conversely, if the transition is driven by an electronic instability, the dominant source of the pre-edge contrast must originate from charge fluctuations rather than from lattice distortions alone. In this case, electronic rearrangements provide the primary contribution to the local symmetry breaking probed at the pre-edge, while the lattice responds by following these electronic fluctuations. The corresponding resonant intensity can then be written as

$$\mathrm{I}(\vec{Q}) \approx 0.3(N_c I_e)$$

where $N_c$ denotes the coherence volume of the charge fluctuations. Figure 5 of the main text compares this expected resonant signal with both the experimental data and the estimated thermal diffuse scattering background for different values of $N_c$. A peak width of 2 eV was used to account for the energy dispersion of the relevant *d* orbitals. From this comparison, we find that very short-ranged charge fluctuations, under a few unit cells, would remain undetectable. Our measurements therefore constrain medium- and long-range charge fluctuations, while leaving open the possibility of short-range electronic correlations.

We should note that we cannot discard fluctuations of any coherent volume with a periodicity not set by half order wavevectors.

### 5. Supplementary figures

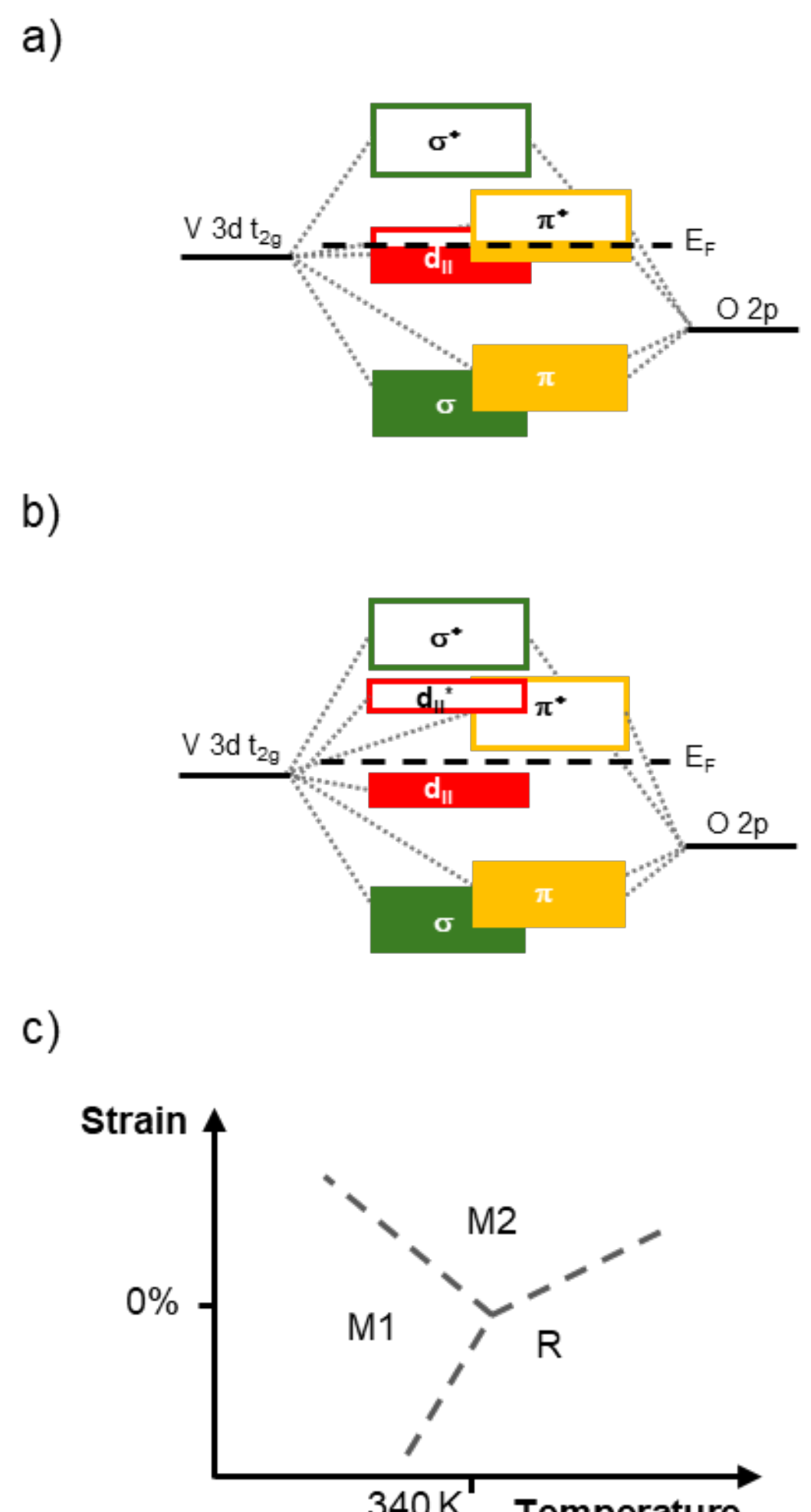

Figure S1. Schematic representation of the $VO_2$ band structure proposed by Goodenough [16] for the a) rutile and b) monoclinic phases. There are three vanadium $t_{2g}$ orbitals. One of them ($d_{||}$) is oriented along the *c*-rutile axis, pointing towards the nearest vanadium atom (See lattice structure in Figure S2). The other two hybridize with the O 2p orbitals, forming a pair of bonding and antibonding $\sigma$ and $\pi$ orbitals. In the metallic rutile phase, both the $d_{||}$ and $\pi^*$ orbitals are crossed by the Fermi surface. There is a slight orbital polarization, with the $d_{||}$ orbital laying lower and having higher occupancy. The structural transition into the monoclinic phase lifts the $\pi^*$orbital, leaving the $d_{||}$ half-filled. Dimerization splits $d_{||}$ into a bonding-antibonding pair, yielding an insulating phase. Unfortunately, the $\sigma$ and $\pi$ notation commonly adopted to designate the $VO_2$ valence orbitals is the same as the one used to distinguish the light polarization in REXS experiments, so we caution the reader to distinguish between both. c) Schematic representation of the $VO_2$ phase diagram as a function of strain and temperature [15].

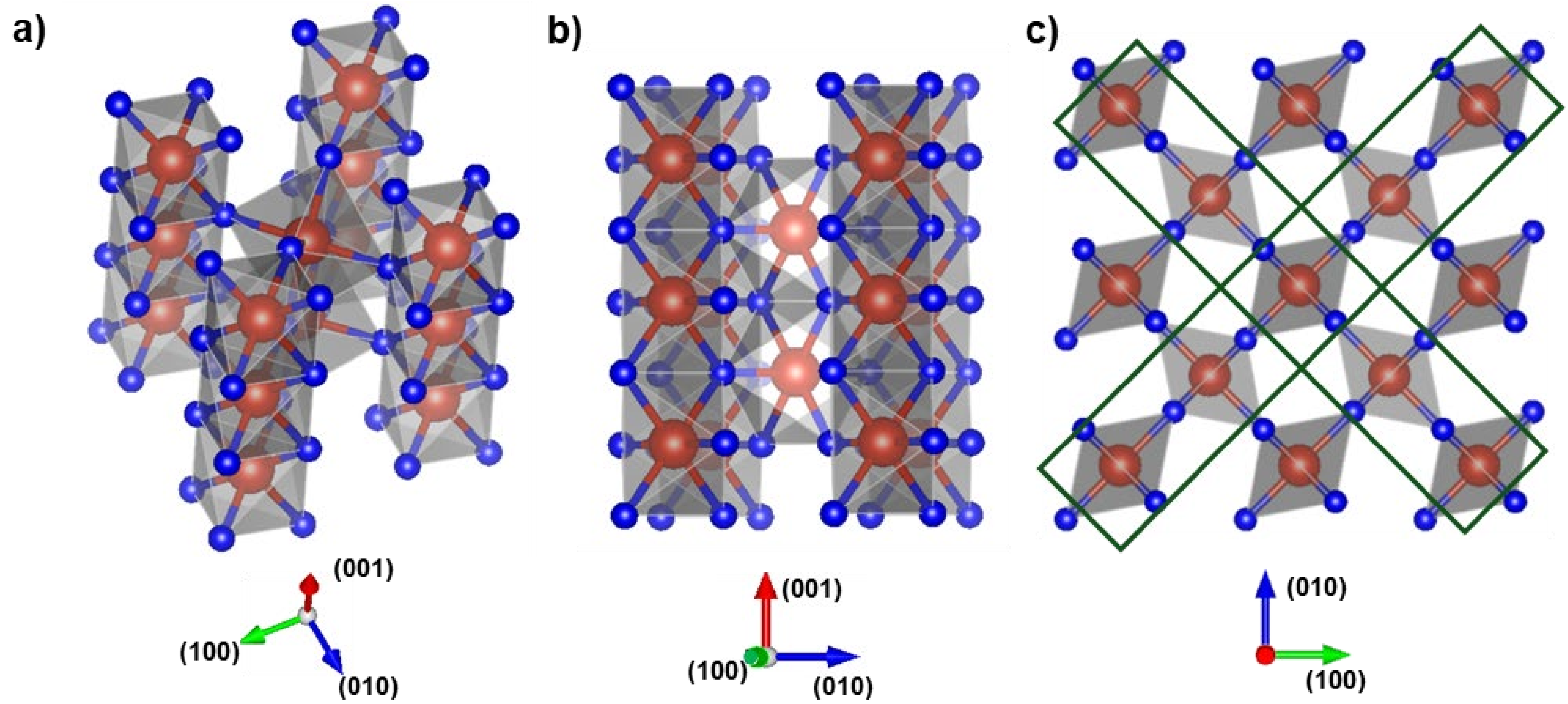

Figure S2. Structure of R $VO_2$, visualized along different directions as indicated by the red/blue/green arrows. Red balls represent vanadium atoms, and blue balls are oxygens. $VO_6$ octahedra are outlined by transparent grey surfaces. The dark green rectangles in c) outline the $(\pm1\pm1\ 0)_R$ planes where structural correlations are strongest [44].

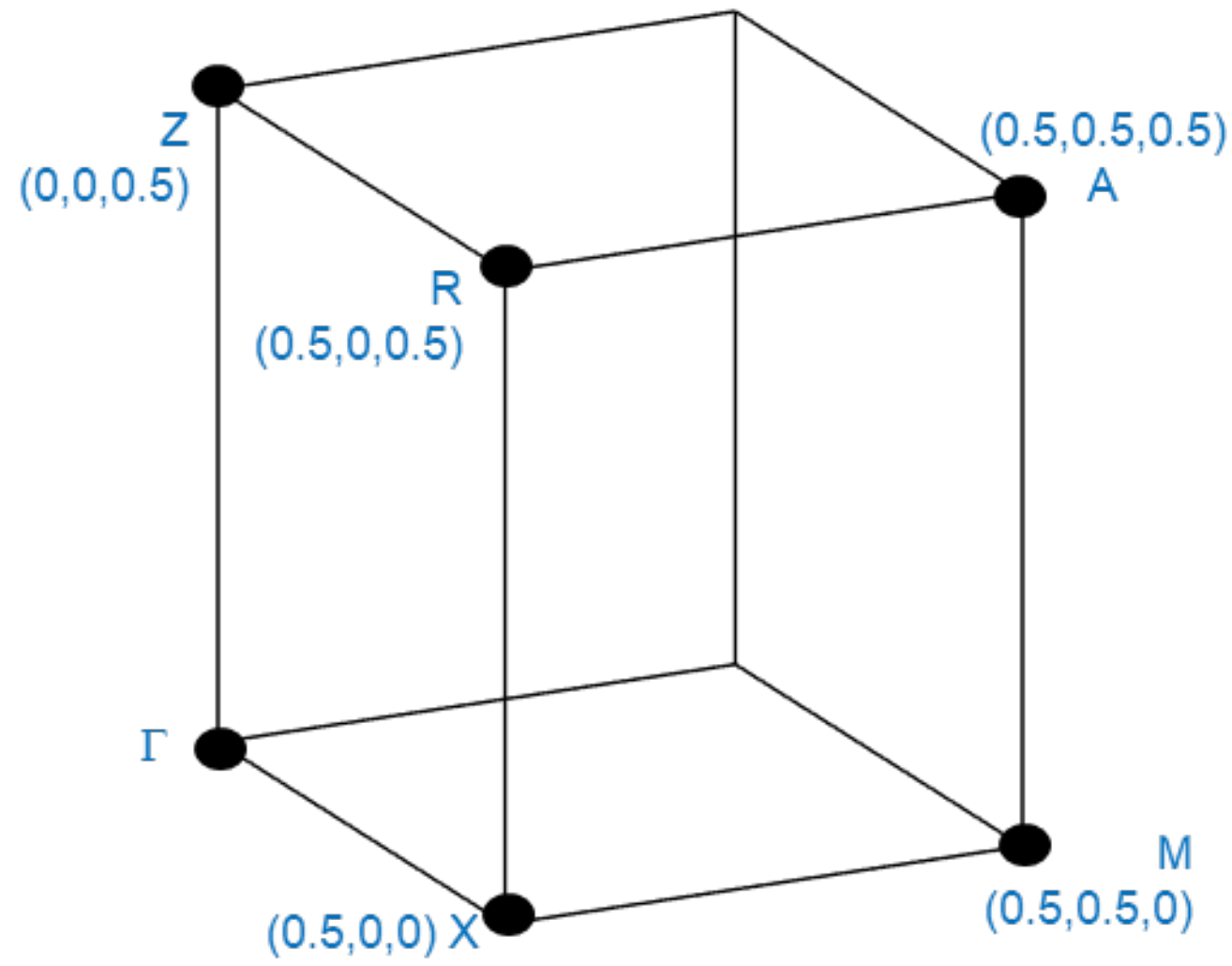

Figure S3. Brillouin zone of the tetragonal rutile structure of the R phase. Black dots show the high symmetry points.

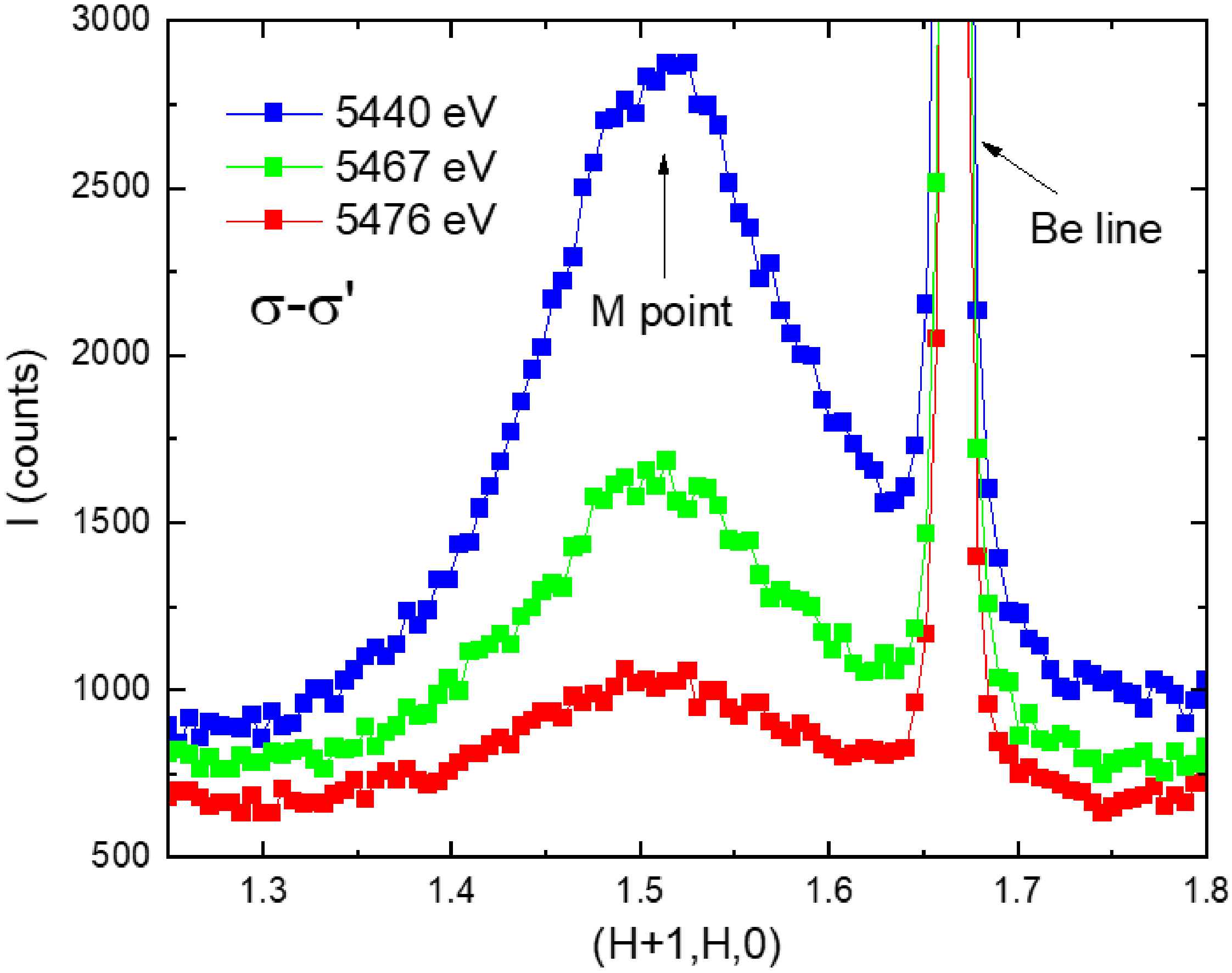

Figure S4. X-ray diffuse scattering along a line scan, following the path outlined with an orange line on the leftmost panel of Figure 3b. Three energies are shown: 30 eV below the absorption edge, at the pre-edge, and at the edge. The sharp feature originates from the Beryllium cryostat dome. The scan intensity has been normalized to better compare the scan shape. Note the large background in this dataset, likely coming from the copper holder.

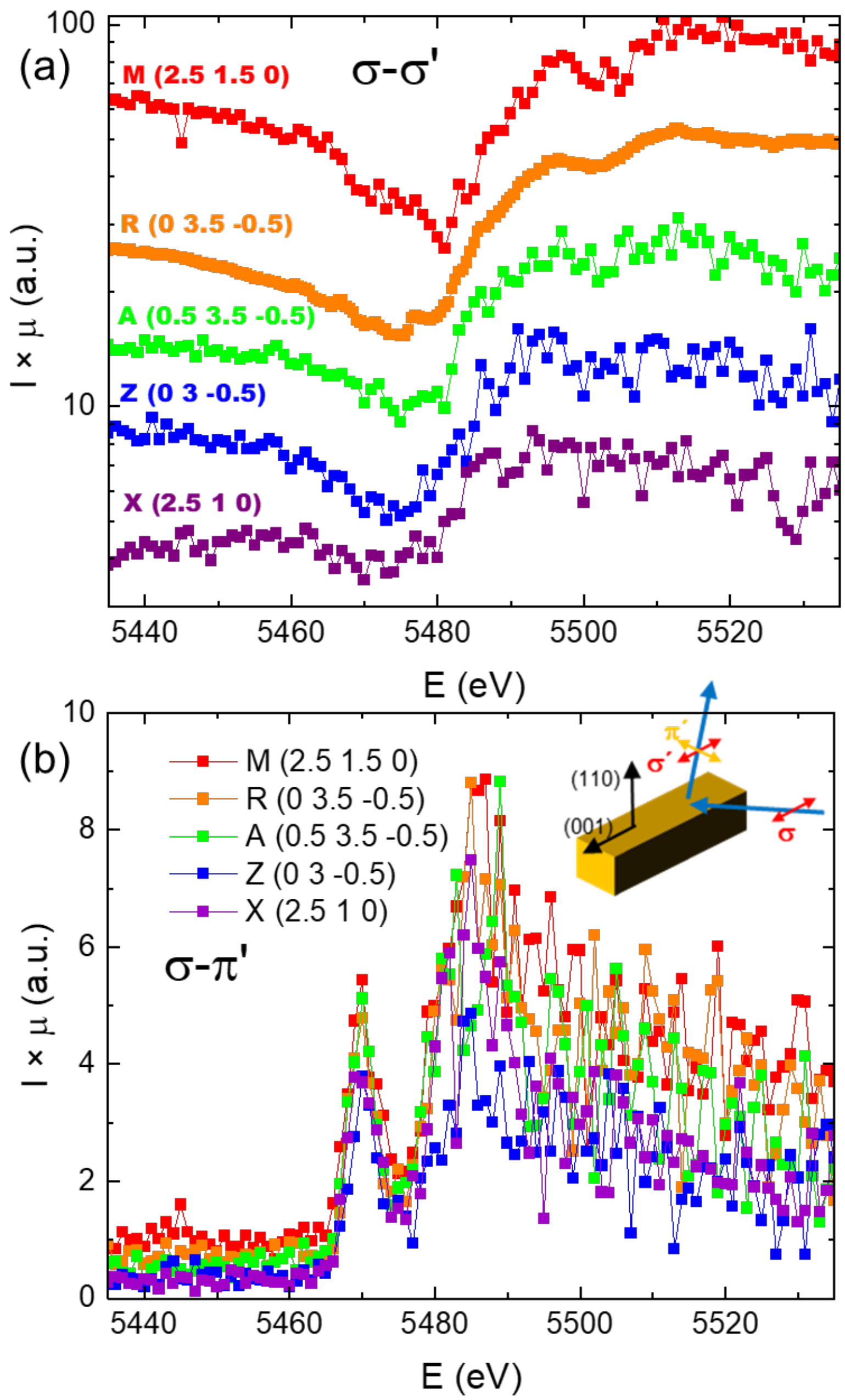

Figure S5. Absorption-corrected X-ray intensity vs photon energy across the K-edge for all high symmetry points of the R structure in: (a) σ-σ´and (b) σ-π´ measurement geometry. T=380 K. Incoming light was polarized nearly parallel to the c-axis, as shown in the schematic drawing.

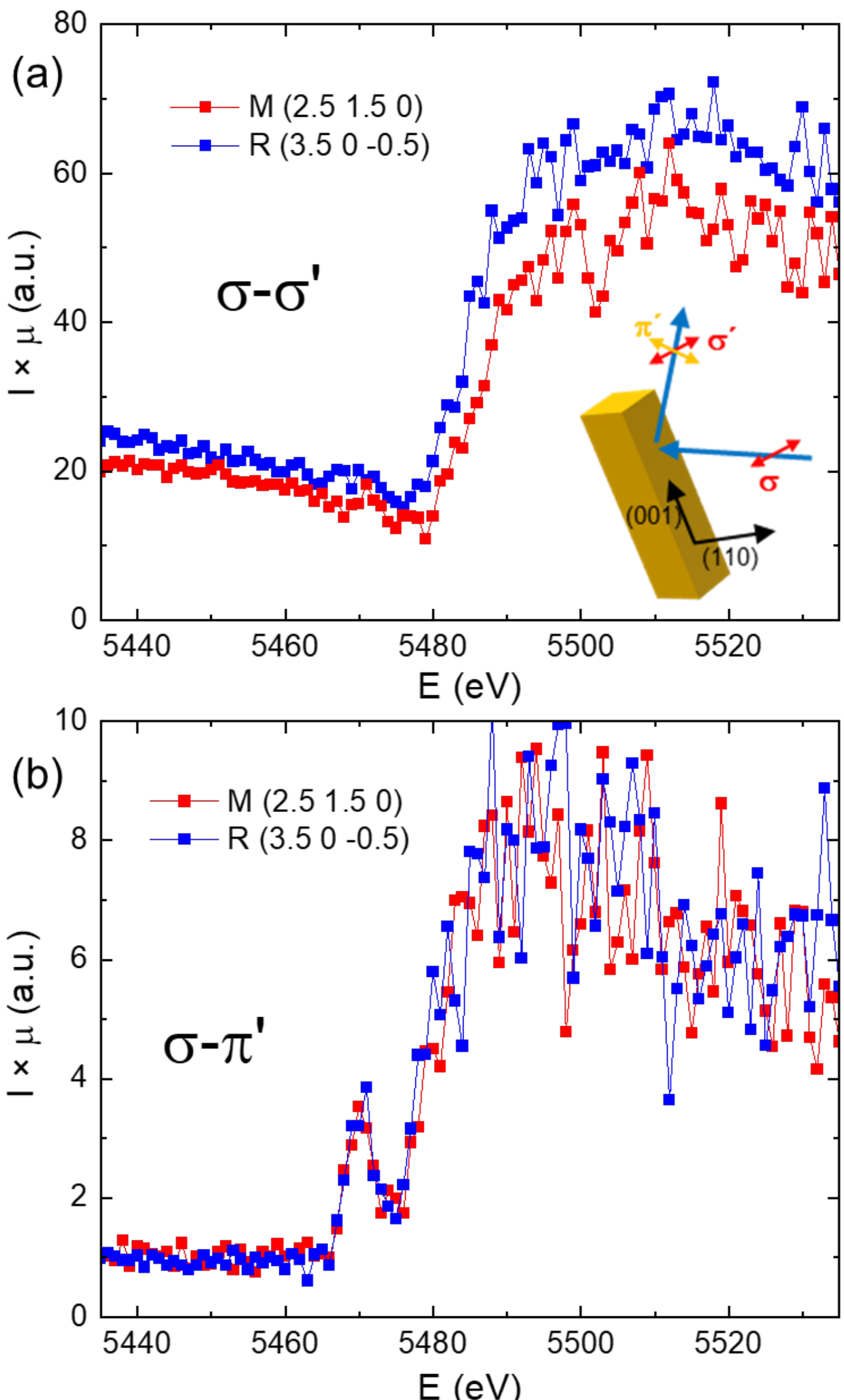

Figure S6. Absorption-corrected X-ray intensity vs photon energy across the K-edge for all high symmetry points of the R structure in: (a) σ-σ´and (b) σ-π´ measurement geometry. T=380 K. Incoming light was polarized nearly perpendicular to the c-axis, as shown in the schematic drawing.

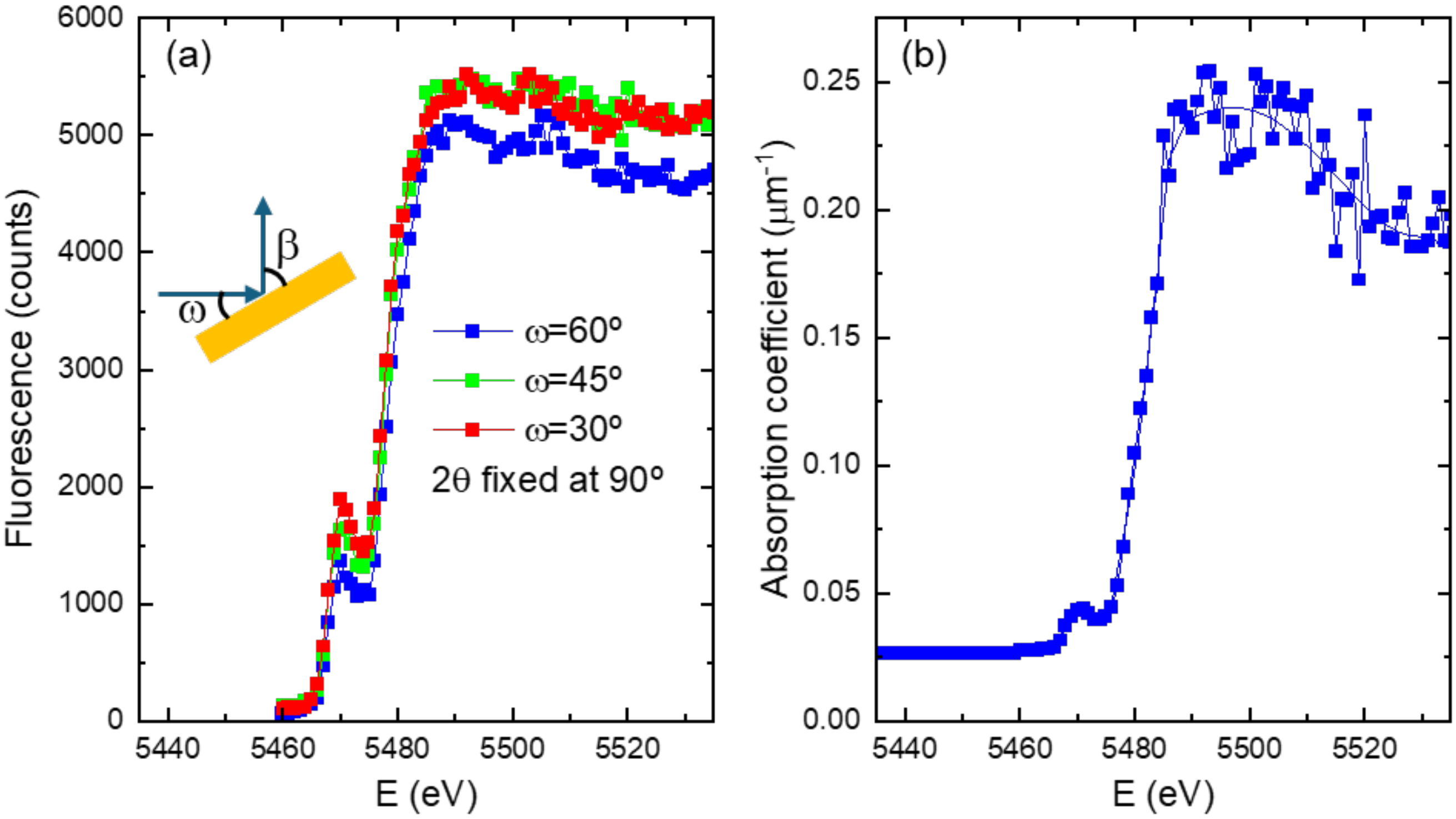

Figure S7. (a) Fluorescence signal vs photon energy for different incidence angles ω. Fluorescence was collected at a fixed 2θ=90º angle. The angle between the surface normal and the scattered light is therefore β=2θ-ω. (b) Absorption coefficient vs photon energy. The solid line shows the data fitting that was used for the calculation of $f'(E)$ and $f''(E)$. T=380 K.

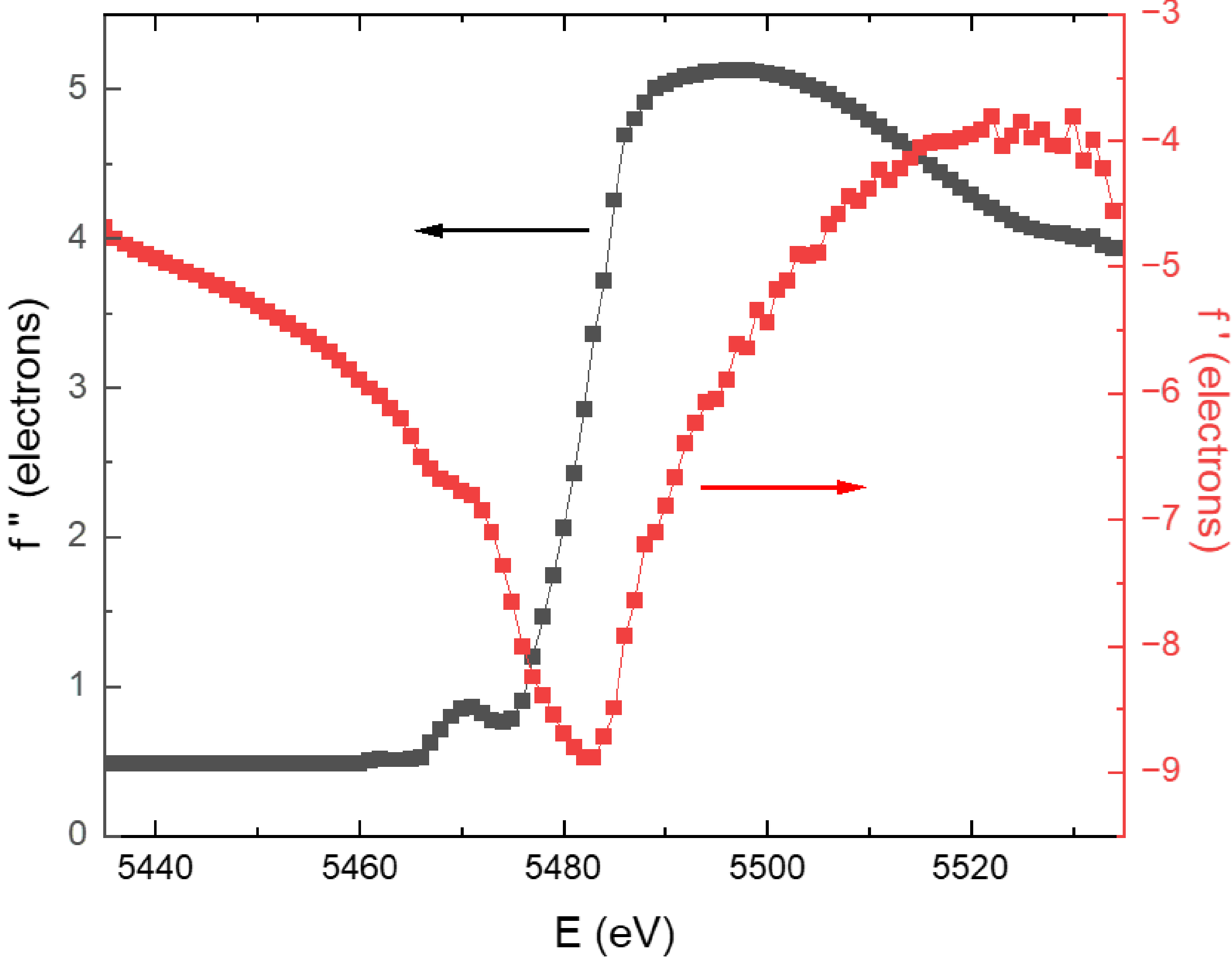

Figure S8. Calculated real ($f'(E)$, red curve) and imaginary ($f''(E)$, black curve) parts of the anomalous scattering factor.

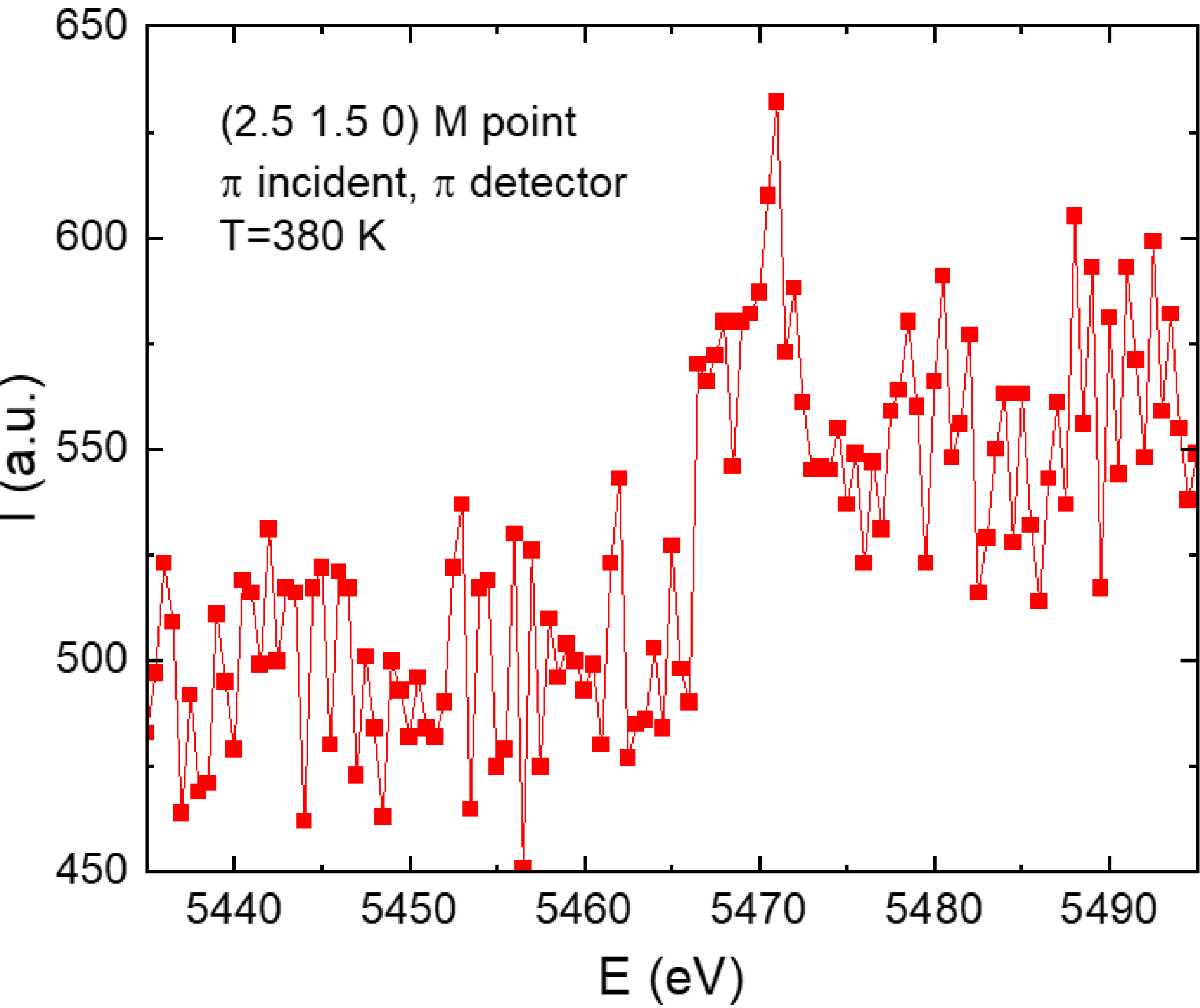

Figure S9. X-ray intensity vs photon energy across the K-edge at the (2.5,1.5,0) M point for a π-π´ measurement geometry. T=380 K. The large background is likely due to spurious scattering from the copper sample holder.

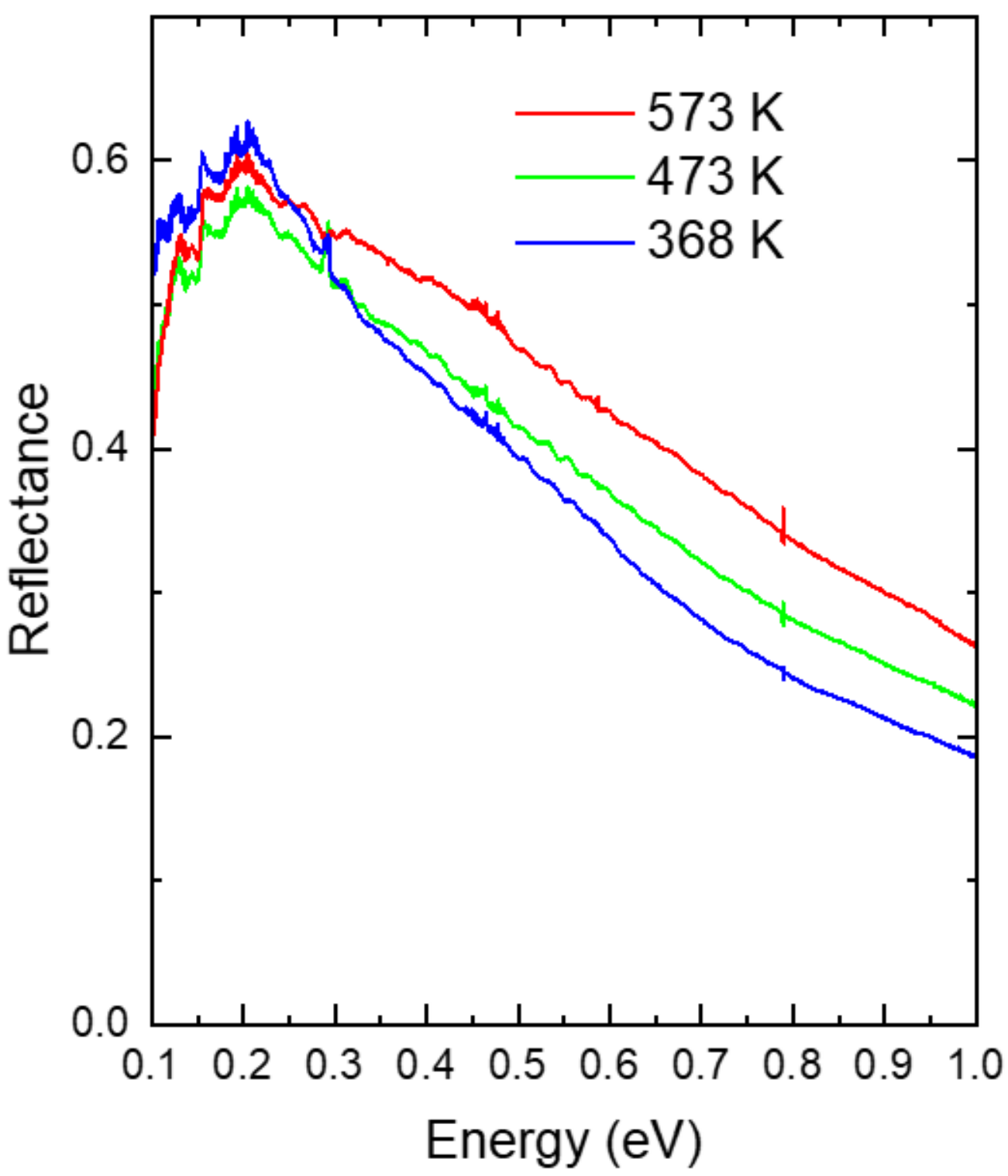

Figure S10. Infrared reflectance vs photon energy, with light polarized along the c axis. Three temperatures above the MIT are shown.

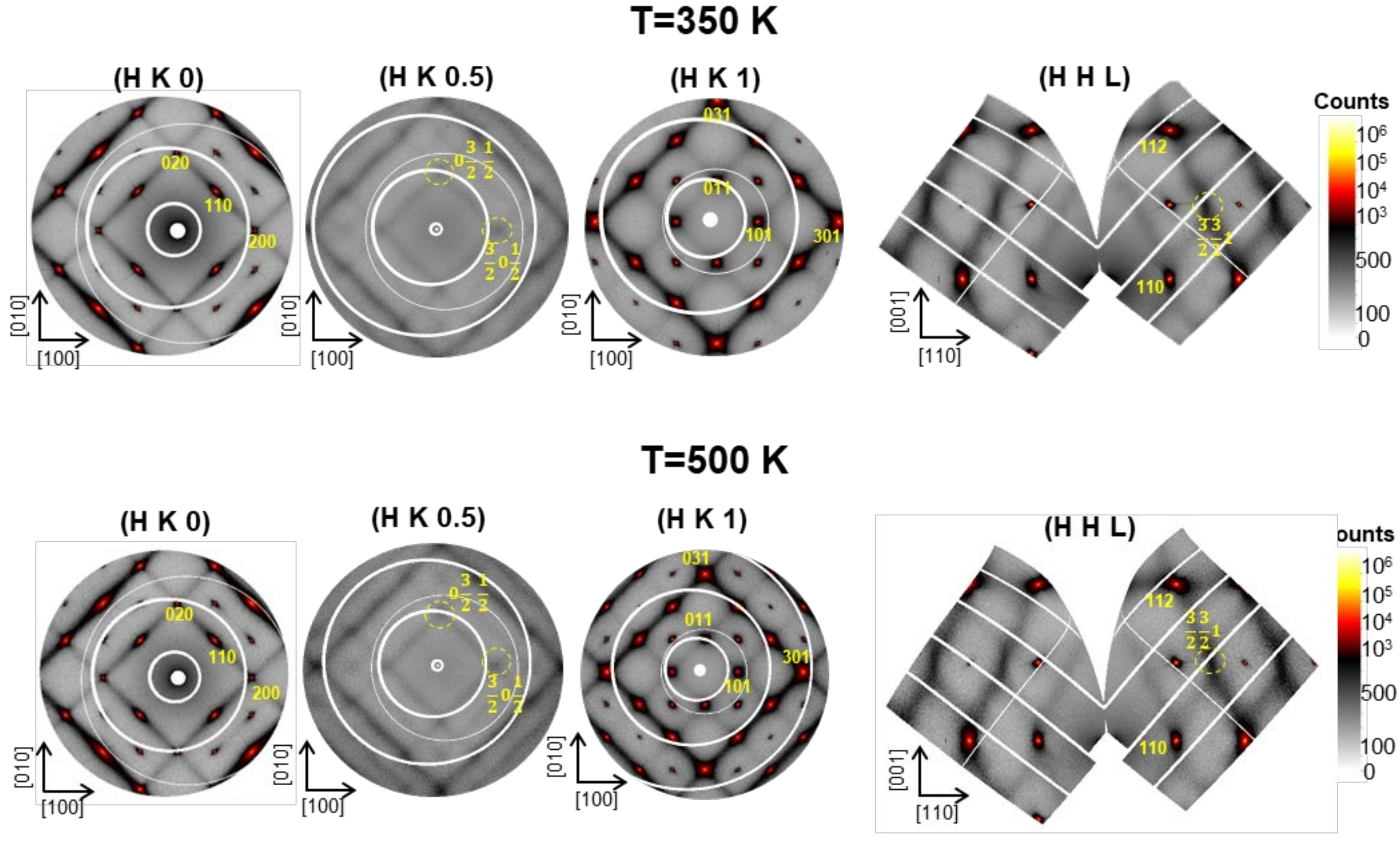

Figure S11. X-ray total scattering in different slices of the reciprocal space: HK0, HK0.5, HK1 and HHL. Two temperatures are shown: 350 K in the upper panels and 500 K in the bottom panels.